\documentclass[titlepage,12pt]{utarticle}
\usepackage{amsmath,amsfonts,bbm,mathrsfs}
\usepackage{epsfig,color}

\input xy
\xyoption{all}

\begin{document}

\date{$1^{st}$ of February 2016}

\title{\vskip-1cm Matrix Factorizations for Local F-Theory Models}
\author{Harun Omer}
\oneaddress{{\tt firstnamelastname@gmail.com}\\{~}\\}

 \nobreak
\Abstract{I use matrix factorizations to describe branes at simple singularities as they appear in elliptic fibrations of local F-theory models.
Each node of the corresponding Dynkin diagrams of the ADE-type singularities is associated with one indecomposable matrix factorization which can be deformed into one or more factorizations of lower rank. Branes with internal fluxes arise naturally as bound states of the indecomposable factorizations. Describing branes in such a way avoids the need to resolve singularities and encodes information which is neglected in conventional F-theory treatments. This paper aims to show how branes arising in local F-theory models  around simple singularities can be described in this framework.}

\maketitle
\pagebreak
\section{Introduction}
Historically, the idea of a category-theoretical description of string theory even predates the discovery of D-branes and first appeared in the context of homological mirror symmetry, where Kontsevich conjectured that the Fukaya category of a Calabi-Yau is equivalent to the derived category of coherent sheaves of the mirror Calabi-Yau~\cite{Kontsevich:1994dn}. Later it was suggested that D-branes can be described as sheaves~\cite{Harvey:1996gc} and brane/anti-brane systems were described as derived categories~\cite{Sharpe:1999qz}. It was in the context of topological boundary Landau-Ginzburg models that D-branes have been explicitly described as and written out as matrix factorizations, using a result of Eisenbud~\cite{Eisenbud}. A selection of papers with particular emphasis on review articles to introduce the subject is found in the
references~\cite{Douglas:2000gi,Diaconescu:2001ze,
Kapustin:2002bi,Kapustin:2003kt,Orlov:2003yp,Sharpe:2003dr,
Lazaroiu:2003md,Brunner:2003dc,Hori:2004zd,Jockers:2007ng}.
\\The category-theoretical description is powerful and its application extends beyond Landau-Ginzburg (LG) models. A strength of the description is its ability to deal with branes at singularities naturally without the need to resolve the geometry. This comes particularly handy in F-theory where 7-branes are located at geometric singularities arising from elliptic
fibrations~\cite{Vafa:1996xn,Donagi:2008ca,Beasley:2008dc,Beasley:2008kw,Donagi:2008kj,
Hayashi:2009ge,Heckman:2009mn,Weigand:2010wm,Katz:2011qp,Callaghan:2011jj,Palti:2012aa,
Camara:2011nj,Maharana:2012tu,Leontaris:2015mpa}.
In particular it has been shown that models with phenomenologically viable features can be built only around a single singularity. This singularity can be thought of as the location of a 7-brane with GUT gauge group, whose singularity type is further enhanced as it intersects other branes at the location of the singularity. In such local F-theory models it is sufficient to focus on the vicinity of the singularity in order to derive physical properties of the theory. Numerous papers have been published in this direction in the last few years.

While 7-branes are no longer D-branes, the general description in terms of sheaves, categories or matrix factorizations is still valid. Collinucci and
Savelli \cite{Collinucci:2014qfa,Collinucci:2014taa} have applied matrix factorization toy models to F-theory. In their work, they sought to rederive a theoretical foundation for applying matrix factorizations to F-theory. Although such a treatment has its own merits, for the purposes of this work, a more pragmatic approach will suffice. We can work with matrix factorizations in F-theory in a similar manner as in Landau-Ginzburg (LG) models. Of course results valid only in the weak-coupling limit such as formulas for topological correlators of the LG models become invalid, but the branes themselves have a description in terms of sheaves and matrix factorizations regardless of the strength of the string coupling and irrespective of whether a part of the target space geometrically specifies the value of the axio-dilaton. Landau-Gizburg (LG) models are valued in a weighted projective target space defined by some equation $W=0$ where $W$ becomes the superpotential of the LG model. The matrix factorizations of $W$ define the branes. LG realizations of matrix factorizations with a torus as target space have been discussed at length, for instance in~\cite{Brunner:2004mt,Govindarajan:2005im,Knapp:2007kq,Omer:2007qs}. The torus can be parameter-dependent and degenerate at certain regions of the parameter space, without affecting the description as matrix factorization. This is essentially what happens
in F-theory models where the torus is elevated to an elliptic fibration with branes located at the singularities. The elliptic fibration can be written in various ways, the simplest of which the Weierstrass equation $y^2=x^3+f(z)x+g(z)$ with $f$ and $g$ appropriate sections. In this paper I will treat this Weierstrass equation (or any alternative description of an elliptic fibration) as the equation of a torus in weighted projective space, so that one can construct a matrix factorization for this torus. To be concrete, we can define,
\begin{eqnarray*}
W(x,y,z) = -y^2 + x^3 +f(z)x+g(z),
\end{eqnarray*}
and treat $W(x,y,z)=0$ as a parameter-dependent torus. The same can be done with other equations of elliptic curves. The matrix factorizations of $W(x,y,z)$ will describe the branes. Matrix factorizations can be parameter-dependent and require no special treatment at points where the parameters take such values that the torus degenerates. This somewhat pragmatic approach contrasts the point of view of Collinucci and Savelli who stress that the application to F-theory is completely distinct from LG models.

With the procedure just explained, one could likewise analyze global F-theory models. Their treatment would however vary on a case-by-case basis. Local models have generic features which can be analyzed, therefore this paper will only deal with local models. I want to move beyond toy models and work with the types of branes which actually appear in phenomenologically viable models. Typically such models start with one GUT\ brane with gauge group $SU(5)$ or $SO(10)$ which intersects other branes.
At the brane intersections, the rank of the gauge group enhances, giving rise to larger symmetry groups. At multiple intersections, enhancements up to $E_8$ are possible. The GUT group can be further broken down by internal fluxes. Particularly relevant for local F-theory models are  the gauge groups $SU(5)$, $SU(6)$ $SO(10)$, $SO(12)$, $E_6$, $E_7$ and $E_8$.
All of these symmetry groups have a geometric description as simple singularities. Simple singularities are classified by the following equations:
\begin{equation}
f(x,y,z)=
\left\{\begin{array}{ll}
-y^2 + x^2 + z^{n+1},     & A_{n}\ \mbox{ with } n\ge 1,\\
-y^2 + x^2z+z^{n-1},& D_n\  \mbox{ with } n\ge 4,\\
-y^2 + x^3+z^4,     & E_6,\\
-y^2 + x^3+xz^3,    & E_7, \\
-y^2 + x^3+z^5,     & E_8\ .
\end{array}\right.\label{eq:simplesing}
\end{equation}
In the main part of this paper the starting point will be the maximal gauge group $E_8$ which is then gradually broken down to smaller subgroups.
\section{Elliptic Fibrations}
\subsection{Equations for Elliptic Curves}
Elliptic curves can be described by different
equations in weighted projective space, such as by a cubic, a quartic or a sextic equation~\cite{Klemm:1996ts}:
\begin{eqnarray}
\begin{array}{l}
x^3 + y^3 + z^3 - a\, x y z \in \mathbbm{P}_2^{1,1,1}\\
x^4 + y^4 + z^2 - a\, x y z \in \mathbbm{P}_2^{1,1,2}\\
x^6 + y^3 + z^2 - a\, x y z \in \mathbbm{P}_2^{1,2,3}
\end{array}
\end{eqnarray} 
An elliptic curve is a nonsingular curve of genus 1 with a rational point.
Although every elliptic curve is topologically equivalent to a torus, different elliptic curves will in general not be isomorphic as Riemann surfaces. 
Isomorphic curves over a field $K$\ have the same $j$-invariant.
Conversely, two curves with the same $j$- invariant are isomorphic over the closure $\bar{K}$. As a consequence of these equivalence theorems, it would be sufficient to consider only one type of equation to describe an elliptic curve. The caveat is that it is the singular fibers of the elliptic fibrations which play the key role in F-theory. Elliptic fibrations which are equivalent as long as they are smooth generically give rise to different types of singularities where the fiber degenerates. It is therefore not sufficient to consider only the standard sextic Weierstrass equation as is done in the vast majority of all research papers in the field.
To more efficiently work with the quartic equation it would be helpful to have a birational transformation between the quartic equation and the standard Weierstrass equation. 
The so-called Tate form is one way to write an elliptic fibration:
\begin{eqnarray}
y^2 + a_1 x y + a_3 y = x^3 + a_2 x^2 + a_4 x + a_6.\label{eq:Tate}
\end{eqnarray}
Each term in the equation can be thought of as being in a graded ring where $x$ has weight 2, $y$ weight $3$ and $a_i$ weight $i$.
By completing the square on the left hand side the $xy$ and $y$ terms can be eliminated after appropriate variable substitution and one obtains,
\begin{eqnarray}
y^2  = x^3 + b_2 x^2 + b_4 x + b_6,\label{eq:bform}
\end{eqnarray}
with,
\begin{eqnarray*}
b_2&=& \frac{1}{4} a_1^2 + a_2 \\
b_4&=&\frac{1}{2}a_1 a_3 + a_4\\ 
b_6&=& \frac{1}{4}a_3^2 + a_6 
\end{eqnarray*}
The advantage of this form for our purposes is its similarity with the equations of simple singularities.
By completing the cube in $x$ the $x^2$-term is eliminated and one obtains the standard Weierstrass equation,
\begin{eqnarray}
y^2  = x^3 + fx+g.\label{eq:Weierstrass}
\end{eqnarray}
These equations in affine form use a local coordinate chart where a point at infinity (the point $z=0$) is left out. This selection of a point defines a global section of the fibration. Alternatively to the Weierstrass equation, an elliptic curve can also be described by a quartic equation.
The quartic equation with general coefficients reads in homogeneous coordinates:
\begin{eqnarray}
v^2 = c_{0}\, u^4 + c_1\, u^3 z + c_2 \,u^2 z^2 + c_3\, u z^3 + c_4\, z^4.\label{eq:quarticHomo}
\end{eqnarray}
By shifting and rescaling coordinates, we can reduce the number of coefficients by two:
\begin{eqnarray}
v^2 = u^4 + c_2\, u^2 z^2 + c_3\, u z^3 + c_4\, z^4.\label{eq:quarticWeier}
\end{eqnarray}
In analogy to the Weierstrass form, we used this to
set the coefficient of the highest order term in $u$ to unity and of the second highest-order term to zero. In affine coordinates the general quartic equation simplifies to,
\begin{eqnarray}
v^2 = c_0 u^4 + c_1 u^3 + c_2 u^2 + c_3 u + c_4.\label{eq:quartic}
\end{eqnarray}
\subsection{A birational transformation between the quartic and the Tate form}
An elliptic curve is an algebraic curve of genus 1 with a rational point on the curve. One point the curve is the point at infinity with projective coordinates $(0 : 1 : 0)$. This rational point with coordinates $(u,v)=(p,q)$ can with the help of a coordinate shift $u$ to $u+p$ be brought into the form $(u,v)=(0,q)$. Since the point must solve Eq.~(\ref{eq:quartic}) we find $c_4=q^2$.
To show birational equivalence to the Weierstrass form, we distinguish between the cases $q = 0$\ and $q\ne 0$.\\\\
The quartic in Eq.~(\ref{eq:quartic}) with $c_4=q^2\ne 0$\ is birationally equivalent to the cubic $y^2 + a_1 x y + a_3 y = x^3 + a_2 x^2 + a_4 x + a_6$ under the transformation~\cite{lwash},
\begin{eqnarray}
\begin{array}{rcl}
x & = &\displaystyle \frac{2 q (v + q)+\ c_3 \, u}{u^2}\\
y &\ = & \displaystyle\frac{2q\left[2q(v+q) + c_3\, u +(c_2-\frac{c_3^2}{4q})u^2\right]}{u^3}
\end{array}
\end{eqnarray}
and the identification,
\begin{eqnarray}
a_1 = \frac{c_3}{q},\;\; \; a_2 = c_2 - \frac{c_3^2}{4 q^2},\; \;\; a_3 = 2 c_1 q,\;\;\; a_4 = -4 c_0 q^2,\; \;\; a_6 = a_4 a_2.
\end{eqnarray}
The equivalence between $(a_1,a_2,a_3,a_4)$\ and $(c_0,c_1,c_2,c_3)$ is one-to-one. In addition we have the constraint 
\begin{eqnarray}
a_6=a_4 a_2 \label{eq:MWrank1}
\end{eqnarray}
on the coefficients of the Weierstrass form as a result of the constraint placed on the quartic curve. The point $(u,v)=(0,q)$ on the curve corresponds to the point at infinity $(x,y)=\infty$ in projective space and the point $(u,v)=(0,-q)$ corresponds to $(x,y)=(-a_2,a_1 a_2 -a_3)$.
The inverse transform is given by,
\begin{eqnarray}
\begin{array}{rcl}
u & = &\displaystyle\frac{2 q \left[x + c_{2} - \frac{c_3^2}{4 q^2}\right]}{y}\\
v &\ = &\displaystyle -q +\frac{-c_3\, u + x u^2}{2q}.
\end{array}
\end{eqnarray}
The above statements are proven by direct calculation, where the equation is multiplied by $2y/u^3$ before the substitution. The argument does not cover the case when the rational point on the curve is at infinity $(u,v)=\infty$. That case implies either $c_{0}=0$ in which case the quartic reduces to a generic cubic, or we have $c_0=q^2 \ne 0$, which allows us to apply the transformation $u\rightarrow 1/u$ and $v\rightarrow v/u^2$ to again obtain an equation of the form in Eq~(\ref{eq:quartic}).\\\\
When $q=0$ the transformation can be written as,
\begin{eqnarray}
\begin{array}{rcl}
u & = &\displaystyle \frac{c_3}{x}\\
v &\ = & \displaystyle\frac{c_3\left[2y+a_1 x+a_3\right]}{2x^2}
\end{array}\label{eq:quarticrational}
\end{eqnarray}
where,
\begin{eqnarray}
c_{0} = \frac{\frac{a_3^2}{4} + a_6}{c_3^2},\,\, c_{1} = \frac{a_1 a_3 + 2 a_4}{2 c_3},\,\,
c_{2} = \frac{a_1^2}{4} + a_2\label{eq:ellipticcoeff}
\end{eqnarray}
The quotients in the variable identification are always defined since  $q=0$ requires $c_3 \ne 0$, otherwise the curve would be singular at the origin $(0,0)$.

 It has been emphasized that the equivalence between birationally equivalent elliptic curves only holds for smooth curves. With the help of Tate's algorithm, the singularities of elliptic curves described by the Tate form has been classified according to the vanishing orders of $a_i$ in the Tate
form. Eq.~(\ref{eq:ellipticcoeff}) is not a valid equation to map the vanishing orders of singularities since the rational transformation between elliptic curves does not necessarily preserve the type of singularity. In~\cite{Kuntzler:2014ila} the Tate algorithm has already been applied to the quartic curve in and a table of singularities was created. In that reference, the general equation is written as,
\begin{eqnarray}
c_0 u^4 + c_1 u^3 z+c_2 u^2 z^2 + c_3 u z^3 +  c_4 z^4 = a v^2 + b_0 z^2 v + b_1 vuz+b_2 u^2 v \label{eq:quarticfull}
\end{eqnarray}
The authors argue that a rank one Mordell-Weil group implies $a=1$, $c_4=0$ and $b_0\ne 0$. By applying these constraints and shifting and scaling the $v$ coordinate, $b_1$ and $b_2$ can be absorbed into new coefficients $c_i$,
\begin{eqnarray}
v^2 = c_0 u^4 + c_1 u^3 z+c_2 u^2 z^2 + c_3 u z^3 + \frac{1}{4} b_0^2 z^4
\end{eqnarray}
Going to affine coordinate by setting $z=1$ we have reproduced Eq.~(\ref{eq:quartic})
with $c_{4}=q^2$ and $q=\displaystyle\frac{ b_0}{2} \ne 0$. One needs to be careful in realizing that Mordell-Weil rank 1 does not necessarily imply that eq.~(\ref{eq:MWrank1})
is always satisfied, only that the elliptic curve can always be brought into a form such that this equation holds. A birational transformation also exists when $c_4$ is not a perfect square, but it is much more complicated and I am not writing it down here.
\subsection{Gauge group breaking of the sextic curve and quartic curves}
In principle one could start with the equation for an elliptic fibration, write out all sections $a_i$ as polynomial expansions and attempt to factorize the equation into a product of two matrices with polynomial entries. To reduce the complexity of the endeavor, we restrict to the main gauge groups of physical interest. The symmetry type determines the vanishing orders of $a_i$ of the Tate form~\cite{Bershadsky:1996nh}:
\begin{eqnarray*}
\begin{array}{ll}
SU(5):&-y^2 + x^3  - a_{1,0} x y\;\;  + a_{2,1} x^2 z\; - a_{3,2} y z^2+ a_{4,3} x z^3  + a_{6,5} z^5\\
SO(10):&-y^2 + x^3  - a_{1,1} x y z  + a_{2,1} x^2 z\; - a_{3,2} y z^2+ a_{4,3} x z^3  + a_{6,5} z^5\\
E_6:&-y^2 + x^3  - a_{1,1} x y z  + a_{2,2} x^2 z^2 - a_{3,2} y z^2+ a_{4,3} x z^3  + a_{6,5} z^5\\
E_7:&-y^2 + x^3  - a_{1,1} x y z  + a_{2,2} x^2 z^2 - a_{3,3} y z^3+ a_{4,3} x z^3  + a_{6,5} z^5\\
E_8:&-y^2 + x^3  - a_{1,1} x y z  + a_{2,2} x^2 z^2 - a_{3,3} y z^3+ a_{4,4} x z^4  + a_{6,5} z^5
\end{array}
\end{eqnarray*}
Here the sections $a_i$ have been expanded into $a_i=\sum_{j}a_{i,j}z^j$.
We will want to deal with all the symmetry groups from $SU(5)$ up to $E_8$ at and therefore wish to preserve all coefficients $a_{i,j}$ which are non-vanishing for any of the groups of interest:
\begin{eqnarray*}
\begin{array}{rcl}
W(x,y,z)&:=&-y^2 + x^3 - a_{1,0} x y - a_{1,1} x y z  + a_{2,1} x^2 z + a_{2,2} x^2 z^2\\
&& - a_{3,2} y z^2- a_{3,3} y z^3 + a_{4,3} x z^3 +a_{4,4} x z^4 + a_{6,5} z^5.
\end{array}
\end{eqnarray*}
We bring this into the $b$-form of eq.~(\ref{eq:bform})
with the transformation,
\begin{eqnarray*}
y \mapsto y - \frac{1}{2}\left(a_{1,0}+a_{1,1}z\right)x-\frac{1}{2}\left(a_{3,2}z^2+a_{3,3}z^3\right)
\end{eqnarray*}
and obtain,
\begin{eqnarray}
\boxed{ W:=-y^2 + f_1 x^3 +f_2 x^2 z+f_3^2 x^2+ 2 f_3 g_3 x z^2+g_1 z^5+g_2 x z^3+g_3^2 z^4}
\label{eq:W}
\end{eqnarray}
where,
\begin{eqnarray*}
\begin{array}{ll}
f_1=1 & g_1=\frac{1}{2}a_{3,2}a_{3,3}+a_{6,5}\\
f_2=\frac{1}{2}a_{1,0}a_{1,1}+a_{2,1} & g_2=\frac{1}{2} a_{1,1} a_{3,2} + \frac{1}{2} a_{1,0} a_{3,3} + a_{4,3}\\
f_3=\frac{1}{2}a_{1,0} & g_3=\frac{1}{2}a_{3,2}
\end{array}
\end{eqnarray*}
In the expression for $W$, three higher order terms have been suppressed: \begin{eqnarray*}
W_{sup}=\left(\frac{1}{4}a_{1,1}^2 + a_{2,2}\right)x^2 z^2+\left(\frac{1}{2} a_{1,1} a_{3,3} + a_{4,4}\right)xz^4+\frac{1}{4}a_{3,3}^2z^6
\end{eqnarray*}
These terms do not affect the singularity type and can therefore be ignored. After all, Eq.~(\ref{eq:W}) is the more fundamental equation with respect to the singularity type and the derivation from the Kodaira classification only serves to relate it to elliptic fibrations.\\

Instead of beginning with the Tate form, we could perform the analogous transformation with the full quartic equation in (\ref{eq:quarticfull}).
In that case one obtains~\cite{Kuntzler:2014ila}:
\begin{eqnarray*}
\begin{array}{ll}
SU(5):&
-y^2 - b_{0,0} x^2 y - b_{1,0} x y\;\; - b_{2,2} yz^2 + c_{0,0} z^5 + c_{1,3} xz^3 + c_{2,1} x^2z\,\; + c_{3,0} x^3\\
SO(10):&-y^2 - b_{0,0} x^2 y - b_{1,1} x y z - b_{2,2} yz^2 + c_{0,0} z^5 + c_{1,3} xz^3 + c_{2,1} x^2z \,\; + c_{3,0} x^3\\
E_6:&-y^2 - b_{0,0} x^2 y - b_{1,1} x y z - b_{2,2} yz^2 + c_{0,0} z^5 + c_{1,3} xz^3 + c_{2,2} x^2z^2 + c_{3,0} x^3\\
E_7:&-y^2 - b_{0,0} x^2 y - b_{1,1} x y z - b_{2,3} yz^3 + c_{0,0} z^5 + c_{1,3} xz^3 + c_{2,2} x^2z^2 + c_{3,0} x^3\\
E_8:&-y^2 - b_{0,0} x^2 y - b_{1,1} x y z- b_{2,3} yz^3 + c_{0,0} z^5 + c_{1,4} xz^4 + c_{2,2} x^2z^2 + c_{3,0} x^3
\end{array}
\end{eqnarray*}
The sections $b_i$ and $c_i$ have been expanded in the obvious manner. The equation with all non-vanishing coefficients preserved reads,
\begin{eqnarray*}
\begin{array}{rcl}
W(x,y,z)&:=& -y^2 - b_{0,0} x^2 y - b_{1,0} x y-b_{1,1}xyz - b_{2,2} yz^2-b_{2,3}yz^3\\&& + c_{0,5}z^5 + c_{1,3} xz^3 + c_{1,4} xz^4 + c_{2,1} x^2z + c_{2,2} x^2z^2 +c_{3,0} x^3
\end{array}
\end{eqnarray*}
After completing the square the resulting equations again has the structure of eq.~(\ref{eq:W}). This time the coefficients mapped as follows:
\begin{eqnarray*}
\begin{array}{ll}
f_1=\frac{1}{2} b_{0,0} b_{1,0} + c_{3,0} & g_1=\frac{1}{2}b_{2,3}b_{2,2}+c_{0,5}\\
f_2=\frac{1}{2}b_{1,0}b_{1,1}+c_{2,1} & g_2=\frac{1}{2} b_{1,1} b_{2,2} + \frac{1}{2} b_{1,0} b_{2,3} + c_{1,3}\\
f_3=\frac{1}{2}b_{1,0} & g_3=\frac{1}{2}b_{2,2}\\
\end{array}
\end{eqnarray*}
Again irrelevant higher order terms have been suppressed:
\begin{eqnarray*}
\begin{array}{rcl}
W_{sup}&=&\frac{1}{4}b_{0,0}^2x^4 + \frac{1}{2} b_{0,0} b_{1,1} x^3 z + \frac{1}{2} b_{0,0} b_{2,3} x^2 z^3 + (\frac{1}{4} b_{1,1}^2 + \frac{1}{2}b_{0,0} b_{2,2}+c_{2,2})x^2 z^2 \\
&&+ (\frac{1}{2} b_{1,1} b_{2,3} + c_{1,4})xz^4 + \frac{1}{4}b_{2,3}^2z^6
\end{array}
\end{eqnarray*}
Eq.~(\ref{eq:W}) will be the starting point of the matrix-factorization based analysis.
\section{Review of Matrix Factorizations}
Given a polynomial $W$ of some coordinate ring in an affine space, a matrix factorization of $W$ is a square matrix $Q$ with polynomial entries so that,
\begin{eqnarray}
Q^2 = W \mathbbm{1}.
\end{eqnarray}
The matrix is in $\mathbbm{Z}_2$-graded space and with the grading operator in suitable form, the matrix can be written as,
\begin{eqnarray}
Q= \left( \begin{array}{cc}0 & E\\ J & 0 \end{array}\right),
\end{eqnarray}
so that,
\begin{eqnarray}
EJ=JE=W\mathbbm{1}.
\end{eqnarray}
Both $Q$ or the pair $(E,J)$ may be referred to as matrix factorization.
The $Q$-notation is rooted in the topological string and the relationship to the boundary $Q_{BRST}$ operator. The simplest factorization is the trivial $1\times 1$ factorization,
\begin{eqnarray}
(1)(W)=(W)(1)=W\mathbbm{1}.
\end{eqnarray}
This trivial factorization describes an 'empty' brane and does not contain physical information. Two matrix factorizations $Q$ and $Q'$ are equivalent, $Q\simeq Q'$, if they can be related by a similarity transformation by invertible matrices with polynomial entries,
\begin{eqnarray}
Q'=U Q U^{-1}.
\end{eqnarray}
The transformation can also be written as,
\begin{eqnarray}
E'=U_1 E U_2^{-1} \qquad J'=U_2 J U_1^{-1},
\end{eqnarray}
with,
\begin{eqnarray}
U= \left(\begin{array}{cc}U_1 & 0\\ 0 & U_2 \end{array}\right).
\end{eqnarray}
Given two matrix factorizations of $W$ we can define the direct sum which is again a matrix factorization of $W$. We have,
\begin{eqnarray}
(E_1,J_1) \oplus (E_2,J_2) \equiv\left( \left( \begin{array}{cccc} E_1 & 0\\
 0 & E_2 \end{array}\right),
 \left( \begin{array}{cccc} J_1 & 0\\
 0 & J_2 \end{array}\right)\right).
\end{eqnarray}
With  the help of  a similarity transformation, some matrix factorizations can be decomposed into a direct sum of factorizations. Given a brane, its anti-brane is easily found by applying the shift functor $T$\ which swaps $E$ and $J$,
\begin{eqnarray}
T: (E,J) \mapsto(\bar{E},\bar{J})=(J,E),
\end{eqnarray}The matrix factorizations $Q_{1,2}$ define a graded differential that acts as follows:
\begin{eqnarray}
d \Psi_{12}:=Q_1\Psi_{12}-(-1)^{|\Psi_{12}|}\Psi_{12}Q_2.
\end{eqnarray}
The open string states lie in the cohomology, which is defined as usual as the quotient of the kernel of $d$\ by the image of $d$. The even states are block diagonal on the $\mathbbm{Z}_2$graded space and are interpreted as bosonic states. The odd states are block off-diagonal and are interpreted as fermions. Specifically, for the fermions we have,
\begin{eqnarray}
d \psi_{12}=\left( \begin{array}{cc}0 & E_1\\ J_1 & 0 \end{array}\right)
\left( \begin{array}{cc}0 & \psi^0_{12}\\ \psi^1_{12} & 0 \end{array}\right)
+\left( \begin{array}{cc}0 & \psi^0_{12}\\ \psi^1_{12} & 0 \end{array}\right)
\left( \begin{array}{cc}0 & E_2\\ J_2 & 0 \end{array}\right),
\end{eqnarray}
and for the bosons,
\begin{eqnarray}
d \phi_{12}=\left( \begin{array}{cc}0 & E_1\\ J_1 & 0 \end{array}\right)
\left( \begin{array}{cc} \phi^0_{12}&0\\ 0 &\phi^1_{12}  \end{array}\right)
-\left( \begin{array}{cc} \phi^0_{12}&0\\ 0 &\phi^1_{12}  \end{array}\right)
\left( \begin{array}{cc}0 & E_2\\ J_2 & 0 \end{array}\right).
\end{eqnarray}
These states can be used to build new factorizations through tachyon condensation.
In tachyon condensation can be described by a short exact sequence of modules,
\begin{eqnarray*}
0 \rightarrow Q_1\rightarrow Q_c \rightarrow Q_2\rightarrow 0.
\end{eqnarray*}
The bound states resulting from tachyon condensation correspond to branes with a flux turned on. In~\cite{Omer:2008fz} it has been shown how non-trivial branes in an orbifold limit of the K3 can be described by matrix factorizations. Fractional branes, discrete Wilson lines as well as unusual orientifold actions all arise from the framework almost automatically.\\\\
Factorizations which can neither be obtained from other factorization by tachyon condensation nor are equivalent to a direct sum of branes are called indecomposable matrix factorization.
For a complete description of all branes on some space described by $W=0$ it is reasonable to begin by finding all indecomposable matrix factorizations of $W$.
For a more detailed and rigorous introduction I refer to the literature cited in the introduction.
 
\section{Simple singularities}
Simple singularities are defined by the following equations:
\begin{equation*}
f(x,y,z)=
\left\{\begin{array}{ll}
-y^2 + x^2 + z^{n+1},     & A_{n}\ \mbox{ with } n\ge 1,\\
-y^2 + x^2z+z^{n-1},& D_n\  \mbox{ with } n\ge 4,\\
-y^2 + x^3+z^4,     & E_6,\\
-y^2 + x^3+xz^3,    & E_7, \\
-y^2 + x^3+z^5,     & E_8\ .
\end{array}\right.
\end{equation*}
The indecomposable matrix factorizations for these ADE singularities are known in the mathematics literature and are given for example in~\cite{Takahashi:2005qu,Kajiura:2005yu}. In the following I list all indecomposable factorizations in a gauge  more suitable for the purposes of this paper. These sets of factorizations should be regarded as the elementary building blocks which through tachyon condensation can fuse into bound states. The bound states obtained in such a fashion correspond  to branes with a non-trivial flux turned on. The indecomposable factorizations are also relevant for global models since in the vicinity of a simple singularity in a global F-theory model, the factorization of the global surface will locally take the same form as these factorizations.
In practice, we will usually work with birational extensions of singularities.
For instance the singularity $y^2=x^3+z^4+z^5$ is an extension
of the $E_6$ singularity $y^2=x^3+z^4$. Results from homological
algebra prove that the morphism between the modules are identical for both singularities.
From the factorizations, the morphisms between them can be found. For the simple singularities this is a solved problem and can be performed with existing computer algebra systems such as 'Singular'~\cite{singular}.
Given a set of matrix factorizations, a quiver diagram can be drawn. Essentially for each distinct irreducible morphism between two factorizations one draws an arrow between them to obtain the so-called Auslander-Reiten quiver. For a proper treatment of them see for example~\cite{yoshino}. In practice, one sets up short exact sequences and reads off the quiver diagram from them. For the ADE-singularities the quivers are essentially the Dynkin diagrams.
To each node in the Dynkin diagram of the corresponding ADE Lie Group corresponds one of the indecomposable factorizations. The rank of the factorization is identical to the degree of the irreducible representation of the Lie Algebra.
For example for the $E_6$ singularity we have six indecomposable factorizations and the short exact sequences,
\begin{eqnarray*}
\begin{array}{cccccccc}
0 &\longrightarrow &M_1&\longrightarrow& M_2 \oplus W &\longrightarrow &M_1&\longrightarrow 0\\
0 &\longrightarrow &M_2&\longrightarrow& M_1\oplus M_3 \oplus M_4& \longrightarrow &M_2&\longrightarrow 0\\
0 &\longrightarrow &M_3&\longrightarrow& M_2\oplus M_5 & \longrightarrow &M_3&\longrightarrow 0\\
0 &\longrightarrow &M_4&\longrightarrow& M_2\oplus M_6 & \longrightarrow &M_4&\longrightarrow 0\\
0 &\longrightarrow &M_5&\longrightarrow& M_3& \longrightarrow &M_5&\longrightarrow 0\\
0 &\longrightarrow &M_6&\longrightarrow& M_4& \longrightarrow &M_6&\longrightarrow 0\\
\end{array}
\end{eqnarray*}
where $W$ stands for the trivial factorization. From these exact sequences we can immediately set up the quiver diagram,
\begin{equation*}
\xymatrix{
 & & \ [W] 
 \ar@<0.7ex>[d] & \\
 & & \ [M^1] 
 \ar@<0.7ex>[d] \ar@<0.7ex>[u] & \\
 \ [M^5]\ \ar@<0.5ex>[r]
 & \ [M^3] \ \ar@<0.5ex>[r]
 \ar@<0.5ex>[l]
 & \ [M^2] \ \ar@<0.5ex>[r]
 \ar@<0.5ex>[l] \ar@<0.5ex>[u]
 & \ [M^4] \ \ar@<0.5ex>[r]
 \ar@<0.5ex>[l]
 & \ [M^6] \ \ . \ar@<0.5ex>[l]
}
\end{equation*}
The trivial factorization $(1,W)$ always corresponds to the extended node of the Dynkin diagram. In case of a singularity which birationally dominates a simple singularity, the quiver diagram will be
identical to the quiver diagram of the dominated singularity except that
the extended node with the trivial factorization is removed
along with all arrows into and out of it. Greuel and Kn\"orrer proved that a ring $R$ is of finite representation type if and only if $R$ birationally dominates a simple singularity~\cite{greuelpfister1985}.
This is a rather strong statement. Finite representation types means that the Auslander-Reiten quiver is of finite size, which we must require for a physically sensible model. This restricts us to simple singularities only.
\subsection{$A_n$ factorizations}
Let,
\begin{eqnarray*}
S_j:=\left(
\begin{array}{cc}
 -x & z^j \\
 z^{n+1-j} & x \\
\end{array}
\right)
\end{eqnarray*}
Then the factorizations of the $A_n$ series singularity are given by,
\begin{eqnarray*}
M^j:(S_j-y\mathbbm{1},S_j+y\mathbbm{1})\qquad j=1,...,n.
\end{eqnarray*}
The quiver diagram can be shown to be,
\begin{equation*}
\xymatrix{
\ \ [M^1]\ \ar@<0.5ex>[r]
& \ [M^2]\ \ar@<0.5ex>[l]\ar@<0.5ex>[r]
& \ \cdot\cdots\cdot\ \ar@<0.5ex>[l]\ar@<0.5ex>[r]
& \ [M^{n-1}]\ \ar@<0.5ex>[l]\ar@<0.5ex>[r]
 & \ [M^n]\ \ar@<0.5ex>[l].
}
\end{equation*}
The factorizations could also be defined for $j=0$ and $j=n+1$, but for these values they reduce to a direct sum of trivial factorizations after a similarity transformation.
The shift functor flips the quiver diagram,
\begin{eqnarray*}
T(M^j)\simeq M^{n+2-j}.
\end{eqnarray*}
\subsection{$D_n$ factorizations}
Let,
\begin{eqnarray*}
S_1:=\left(
\begin{array}{cc}
 0 & x^2 + z^{n-2} \\
 z & 0 \\
\end{array}
\right),
\end{eqnarray*}
\begin{eqnarray*}
S_j:=\left(
\begin{array}{cccc}
 0 & 0 & x z & z^{n-1-\frac{j}{2}} \\
 0 & 0 & z^{\frac{j}{2}} & -x \\
 x & z^{n-1-\frac{j}{2}} & 0 & 0 \\
 z^{\frac{j}{2}} & -x z & 0 & 0 \\
\end{array}
\right)\;\text{ for }j \text{ even and } 2 \le j \le n-2,
\end{eqnarray*}
\begin{eqnarray*}
S_j:=\left(
\begin{array}{cccc}
 0 & 0 & x z & z^{n-1+\frac{1-j}{2}} \\
 0 & 0 & z^{\frac{j+1}{2}} & -x z \\
 x & z^{n-2+\frac{1-j}{2}} & 0 & 0 \\
 z^{\frac{j-1}{2}} & -x & 0 & 0 \\
\end{array}
\right)\;\text{ for }j \text{ odd and } 2 \le j \le n-2,
\end{eqnarray*}
\begin{eqnarray*}
S_{n-1}:=\left(
\begin{array}{cc}
 0 & x z-i z^{\frac{n}{2}} \\
 x-i z^{\frac{n}{2}-1} & 0 \\
\end{array}
\right)\text{ when }n \text{ is even},
\end{eqnarray*}
\begin{eqnarray*}
S_{n-1}:=\left(
\begin{array}{cc}
 -z^{\frac{n-1}{2}} & xz \\
 x  & z^{\frac{n-1}{2}} \\
\end{array}
\right)\text{ when }n \text{ is odd},
\end{eqnarray*}
\begin{eqnarray*}
S_{n}:=\left(
\begin{array}{cc}
 0 & x z+i z^{\frac{n}{2}} \\
x-i z^{\frac{n}{2}-1}  & 0 \\
\end{array}
\right)\text{ when }n \text{ is even},
\end{eqnarray*}
\begin{eqnarray*}
S_{n}:=\left(
\begin{array}{cc}
 z^{\frac{n-1}{2}} & xz \\
 x & -z^{\frac{n-1}{2}} \\
\end{array}
\right)\text{ when }n \text{ is odd}.
\end{eqnarray*}
Then the factorizations of the $D_n$ series singularities are given by,
\begin{eqnarray*}
M^j:(S_j-y\mathbbm{1},S_j+y\mathbbm{1})\qquad j=1,...,n.
\end{eqnarray*}
The quiver diagram takes the form,
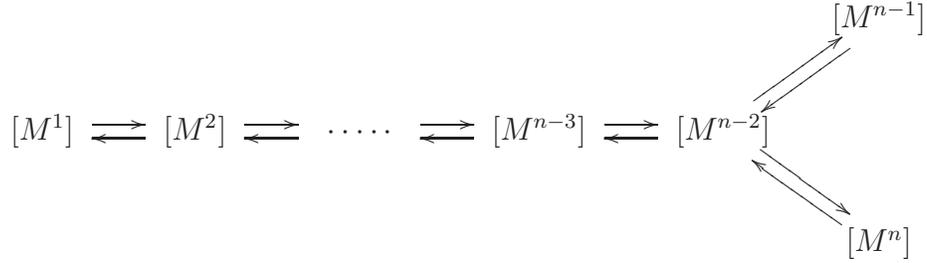
\begin{equation*}
\xymatrix@C=.6em{
 && && && && &\ [M^{n-1}]\  \ar@<0.5ex>[ld]\\
\ [M^1]\ \ar@<0.5ex>[rr]
  && \ [M^2]\ \ar@<0.5ex>[ll]
 \ar@<0.5ex>[rr] 
  && \ \ \cdot\cdots\cdot \ \ \ar@<0.5ex>[ll]
  \ar@<0.5ex>[rr]  
  &&\ [M^{n-3}]\ 
 \ar@<0.5ex>[rr] \ar@<0.5ex>[ll]
  &&\ [M^{n-2}]\ 
 \ar@<0.5ex>[ru] \ar@<0.5ex>[rd]
 \ar@<0.5ex>[ll] \\
 && && && && & 
 \ \ \ [M^n]\ \ar@<0.5ex>[lu] \ \
}
\end{equation*}
All $D_n$ factorizations are self-dual under $T$ except for $M^{n}$ and $M^{n-1}$ which are duals of each other:
\begin{eqnarray*}
\begin{array}{lcl}
T(M^j)&\simeq&M^{j}\text{ for }j=1,...,n-2\\
T(M^{n-1})&\simeq&M^n\\
T(M^{n})&\simeq&M^{n-1}
\end{array}
\end{eqnarray*}
\subsection{$E_6$ factorizations}
Let,
\begin{eqnarray*}
S_1:=\left(
\begin{array}{cccc}
 0 & 0 & x^2 & z^3 \\
 0 & 0 & z & -x \\
 x & z^3 & 0 & 0 \\
 z & -x^2 & 0 & 0 \\
\end{array}
\right),
\end{eqnarray*}
\begin{eqnarray*}
S_2:=\left(
\begin{array}{cccccc}
 0 & 0 & 0 & x^2 & -x z & z^2 \\
 0 & 0 & 0 & z^3 & x^2 & -x z \\
 0 & 0 & 0 & -x z^2 & z^3 & x^2 \\
 x & z & 0 & 0 & 0 & 0 \\
 0 & x & z & 0 & 0 & 0 \\
 z^2 & 0 & x & 0 & 0 & 0 \\
\end{array}
\right),
\end{eqnarray*}
\begin{eqnarray*}
S_3:=\left(
\begin{array}{cccc}
 z^2 & i x^2 & 0 & -x z \\
 -i x & -z^2 & -z & 0 \\
 0 & 0 & z^2 & i x^2 \\
 0 & 0 & -i x & -z^2 \\
\end{array}
\right),
\end{eqnarray*}
\begin{eqnarray*}
S_5:=\left(
\begin{array}{cc}
 z^2 & x^2 \\
 x & -z^2 \\
\end{array}
\right),
\end{eqnarray*}
Then the six indecomposable factorizations of the $E_6$ singularity are given by,
\begin{eqnarray*}
M^1:(S_1-y\mathbbm{1},S_1+y\mathbbm{1})\\
M^2:(S_2-y\mathbbm{1},S_2+y\mathbbm{1})\\
M^3:(S_3-y\mathbbm{1},S_3+y\mathbbm{1})\\
M^4:(S_3+y\mathbbm{1},S_3-y\mathbbm{1})\\
M^5:(S_5-y\mathbbm{1},S_5+y\mathbbm{1})\\
M^6:(S_5+y\mathbbm{1},S_5-y\mathbbm{1})
\end{eqnarray*}
The quiver diagram is,
\begin{equation*}
\xymatrix{
 & & \ [M^1] 
 \ar@<0.7ex>[d] & \\
 \ [M^5]\ \ar@<0.5ex>[r]
 & \ [M^3] \ \ar@<0.5ex>[r]
 \ar@<0.5ex>[l]
 & \ [M^2] \ \ar@<0.5ex>[r]
 \ar@<0.5ex>[l] \ar@<0.5ex>[u]
 & \ [M^4] \ \ar@<0.5ex>[r]
 \ar@<0.5ex>[l]
 & \ [M^6] \ \ . \ar@<0.5ex>[l]
}
\end{equation*}
The actions of the functor $T$ is given by,
\begin{eqnarray*}
\begin{array}{lcl}
T(M^1)&=&M^1\\
T(M^2)&=&M^2\\
T(M^3)&=&M^4\\
T(M^4)&=&M^3\\
T(M^5)&=&M^6\\
T(M^6)&=&M^5
\end{array}
\end{eqnarray*}
\subsection{$E_7$ factorizations}
Let,
\begin{eqnarray*}
S_1:=\left(
\begin{array}{cccc}
 0 & 0 & x^2 & x z^2 \\
 0 & 0 & z & -x \\
 x & x z^2 & 0 & 0 \\
 z & -x^2 & 0 & 0 \\
\end{array}
\right),
\end{eqnarray*}
\begin{eqnarray*}
S_2:=\left(
\begin{array}{cccccc}
 0 & 0 & 0 & x^2 & x z^2 & -x^2 z \\
 0 & 0 & 0 & -x z & x^2 & x z^2 \\
 0 & 0 & 0 & z^2 & -x z & x^2 \\
 x & 0 & x z & 0 & 0 & 0 \\
 z & x & 0 & 0 & 0 & 0 \\
 0 & z & x & 0 & 0 & 0 \\
\end{array}
\right),
\end{eqnarray*}
\begin{eqnarray*}
S_3:=\left(
\begin{array}{cccccccc}
 0 & 0 & 0 & 0 & x & x z & 0 & z \\
 0 & 0 & 0 & 0 & z^2 & -x^2 & -x z & 0 \\
 0 & 0 & 0 & 0 & 0 & 0 & x^2 & z^2 \\
 0 & 0 & 0 & 0 & 0 & 0 & x z & -x \\
 x^2 &  x z & 0 & x z & 0 & 0 & 0 & 0 \\
 z^2 & -x & -z & 0 & 0 & 0 & 0 & 0 \\
 0 & 0 & x & z^2 & 0 & 0 & 0 & 0 \\
 0 & 0 & x z & -x^2 & 0 & 0 & 0 & 0 \\
\end{array}
\right),
\end{eqnarray*}
\begin{eqnarray*}
S_4:=\left(
\begin{array}{cccc}
 0 & 0 & x z & x^2 \\
 0 & 0 & x^2 & -x z^2 \\
 z^2 & x & 0 & 0 \\
 x & -z & 0 & 0 \\
\end{array}
\right),
\end{eqnarray*}
\begin{eqnarray*}
S_5:=\left(
\begin{array}{cccccc}
 0 & 0 & 0 & x^2 & z^2 & -x z \\
 0 & 0 & 0 & -x z & x & z^2 \\
 0 & 0 & 0 & x z^2 & -x z & x^2 \\
 x & 0 & z & 0 & 0 & 0 \\
 x z & x^2 & 0 & 0 & 0 & 0 \\
 0 & x z & x & 0 & 0 & 0 \\
\end{array}
\right),
\end{eqnarray*}
\begin{eqnarray*}
S_6:=\left(
\begin{array}{cccc}
 0 & 0 & x^2 & x z \\
 0 & 0 & z^2 & -x \\
 x & x z & 0 & 0 \\
 z^2 & -x^2 & 0 & 0 \\
\end{array}
\right),
\end{eqnarray*}
\begin{eqnarray*}
S_7:=\left(
\begin{array}{cc}
 0 & z^3+x^2 \\
 x & 0 \\
\end{array}
\right),
\end{eqnarray*}
Then the factorizations of the $E_7$ singularity are given by,
\begin{eqnarray*}
M^j:(S_j-y\mathbbm{1},S_j+y\mathbbm{1})\qquad j=1,...,7.
\end{eqnarray*}
The quiver diagram is,
\begin{equation*}
\xymatrix{
 & & & [M^4] \ar@<0.5ex>[d] & & & \\
 \ [M^7] \  \ar@<0.5ex>[r] 
 & \ [M^6] \  \ar@<0.5ex>[r] 
 \ar@<0.5ex>[l] 
 & \ [M^5] \  
 \ar@<0.5ex>[r] \ar@<0.5ex>[l]
 & \ [M^3] \  \ar@<0.5ex>[r]
 \ar@<0.5ex>[l] \ar@<0.5ex>[u]
 & \ [M^2] \  
\ar@<0.5ex>[r] \ar@<0.5ex>[l]
 & \ [M^1] \ . 
\ar@<0.5ex>[l]
 & 
}
\end{equation*}
All the $E_7$ factorizations are self-dual under $T$.
\subsection{$E_8$ factorizations}
Let,
\begin{eqnarray*}
S_1:=\left(
\begin{array}{cccc}
 0 & 0 & x^2 & z^4 \\
 0 & 0 & z & -x \\
 x & z^4 & 0 & 0 \\
 z & -x^2 & 0 & 0 \\
\end{array}
\right),
\end{eqnarray*}
\begin{eqnarray*}
S_2:=\left(
\begin{array}{cccccc}
 0 & 0 & 0 & x^2 & z^4 & -x z^3 \\
 0 & 0 & 0 & -x z & x^2 & z^4 \\
 0 & 0 & 0 & z^2 & -x z & x^2 \\
 x & 0 & z^3 & 0 & 0 & 0 \\
 z & x & 0 & 0 & 0 & 0 \\
 0 & z & x & 0 & 0 & 0 \\
\end{array}
\right),
\end{eqnarray*}
\begin{eqnarray*}
S_3:=\left(
\begin{array}{cccccccc}
 0 & 0 & 0 & 0 & x & z^3 & 0 & z \\
 0 & 0 & 0 & 0 & z^2 & -x^2 & -x z & 0 \\
 0 & 0 & 0 & 0 & 0 & 0 & x^2 & z^2 \\
 0 & 0 & 0 & 0 & 0 & 0 & z^3 & -x \\
 x^2 & z^3 & 0 & x z & 0 & 0 & 0 & 0 \\
 z^2 & -x & -z & 0 & 0 & 0 & 0 & 0 \\
 0 & 0 & x & z^2 & 0 & 0 & 0 & 0 \\
 0 & 0 & z^3 & -x^2 & 0 & 0 & 0 & 0 \\
\end{array}
\right),
\end{eqnarray*}
\begin{eqnarray*}
S_4:=\left(
\begin{array}{cccccccccc}
 0 & 0 & 0 & 0 & 0 & x^2 & z^2 & z^3 & 0 & -x z \\
 0 & 0 & 0 & 0 & 0 & z^3 & -x & 0 & -z^2 & 0 \\
 0 & 0 & 0 & 0 & 0 & 0 & 0 & x^2 & -x z & z^3 \\
 0 & 0 & 0 & 0 & 0 & 0 & 0 & z^4 & x^2 & -x z^2 \\
 0 & 0 & 0 & 0 & 0 & 0 & 0 & -x z^2 & z^3 & x^2 \\
 x & z^2 & 0 & 0 & z & 0 & 0 & 0 & 0 & 0 \\
 z^3 & -x^2 & 0 & -z^2 & 0 & 0 & 0 & 0 & 0 & 0 \\
 0 & 0 & x & z & 0 & 0 & 0 & 0 & 0 & 0 \\
 0 & 0 & 0 & x & z^2 & 0 & 0 & 0 & 0 & 0 \\
 0 & 0 & z^2 & 0 & x & 0 & 0 & 0 & 0 & 0 \\
\end{array}
\right),
\end{eqnarray*}
\begin{eqnarray*}
S_5:=\left(
\begin{array}{cccccccccccc}
 0 & 0 & 0 & 0 & 0 & 0 & x^2 & -x z & z^3 & 0 & x & 0 \\
 0 & 0 & 0 & 0 & 0 & 0 & z^4 & x^2 & -x z^2 & 0 & 0 & x z \\
 0 & 0 & 0 & 0 & 0 & 0 & -x z^2 & z^3 & x^2 & x z & 0 & 0 \\
 0 & 0 & 0 & 0 & 0 & 0 & 0 & 0 & 0 & x & z & 0 \\
 0 & 0 & 0 & 0 & 0 & 0 & 0 & 0 & 0 & 0 & x & z^2 \\
 0 & 0 & 0 & 0 & 0 & 0 & 0 & 0 & 0 & z^2 & 0 & x \\
 x & z & 0 & 0 & -x & 0 & 0 & 0 & 0 & 0 & 0 & 0 \\
 0 & x & z^2 & 0 & 0 & -x z & 0 & 0 & 0 & 0 & 0 & 0 \\
 z^2 & 0 & x & -x z & 0 & 0 & 0 & 0 & 0 & 0 & 0 & 0 \\
 0 & 0 & 0 & x^2 & -x z & z^3 & 0 & 0 & 0 & 0 & 0 & 0 \\
 0 & 0 & 0 & z^4 & x^2 & -x z^2 & 0 & 0 & 0 & 0 & 0 & 0 \\
 0 & 0 & 0 & -x z^2 & z^3 & x^2 & 0 & 0 & 0 & 0 & 0 & 0 \\
\end{array}
\right),
\end{eqnarray*}
\begin{eqnarray*}
S_6:=\left(
\begin{array}{cccccc}
 0 & 0 & 0 & x^2 & -x z & z^3 \\
 0 & 0 & 0 & z^4 & x^2 & -x z^2 \\
 0 & 0 & 0 & -x z^2 & z^3 & x^2 \\
 x & z & 0 & 0 & 0 & 0 \\
 0 & x & z^2 & 0 & 0 & 0 \\
 z^2 & 0 & x & 0 & 0 & 0 \\
\end{array}
\right),
\end{eqnarray*}
\begin{eqnarray*}
S_7:=\left(
\begin{array}{cccccccc}
 0 & 0 & 0 & 0 & x & z^2 & 0 & -z \\
 0 & 0 & 0 & 0 & z^3 & -x^2 & x z^2 & 0 \\
 0 & 0 & 0 & 0 & 0 & 0 & x^2 & z^2 \\
 0 & 0 & 0 & 0 & 0 & 0 & z^3 & -x \\
 x^2 & z^2 & 0 & -x z & 0 & 0 & 0 & 0 \\
 z^3 & -x & z^2 & 0 & 0 & 0 & 0 & 0 \\
 0 & 0 & x & z^2 & 0 & 0 & 0 & 0 \\
 0 & 0 & z^3 & -x^2 & 0 & 0 & 0 & 0 \\
\end{array}
\right),
\end{eqnarray*}
\begin{eqnarray*}
S_8:=\left(
\begin{array}{cccc}
 0 & 0 & x^2 & z^3 \\
 0 & 0 & z^2 & -x \\
 x & z^3 & 0 & 0 \\
 z^2 & -x^2 & 0 & 0 \\
\end{array}
\right),
\end{eqnarray*}
Then the factorizations of the $E_8$ singularity are given by,
\begin{eqnarray*}
M^j:(S_j-y\mathbbm{1},S_j+y\mathbbm{1})\qquad j=1,...,8.
\end{eqnarray*}
The quiver diagram is,
\begin{equation*}
\hspace*{-0.7cm}
\xymatrix{
 & & & & [M^6] \ar@<0.5ex>[d]^{} & & \\
 \ [M^1] \  \ar@<0.5ex>[r]^{} 
 & \ [M^2] \  
 \ar@<0.5ex>[r]^{} \ar@<0.5ex>[l]^{} 
 & \ [M^3] \  
 \ar@<0.5ex>[r]^{} \ar@<0.5ex>[l]^{} 
 & \ [M^4] \  
\ar@<0.5ex>[r]^{} \ar@<0.5ex>[l]^{} 
 & \ [M^5] \  
\ar@<0.5ex>[r]^{} \ar@<0.5ex>[l]^{} \ar@<0.5ex>[u]^{}
 & \ [M^7] \  
\ar@<0.5ex>[r]^{} \ar@<0.5ex>[l]^{} 
 & \ [M^8] \ . 
\ar@<0.5ex>[l]^{} 
}
\end{equation*}
All the $E_7$ factorizations are self-dual under $T$.
\section{Deformations of Matrix Factorizations}
The matrix factorizations for the $A_n$ surface singularity were listed above as,
\begin{eqnarray*}
E_j=\left(
\begin{array}{cc}
 -x-y & z^j \\
 z^{n+1-j} & x-y \\
\end{array}
\right)\qquad
J_j=\left(
\begin{array}{cc}
 -x+y & z^j \\
 z^{n+1-j} & x+y \\
\end{array}
\right)\qquad j=1,...,n.
\end{eqnarray*}
To be concrete, I set $n=5$ and select the factorization with $j=2$ so that,
\begin{eqnarray}
E_2 \cdot J_2 = 
\left(
\begin{array}{cc}
 -x-y & z^2 \\
 z^4 & x-y \\
\end{array}
\right)
\cdot
\left(
\begin{array}{cc}
 -x+y & z^2 \\
 z^4 & x+y \\
\end{array}
\right)
=(-y^2+x^2+z^6)\mathbbm{1}.
\end{eqnarray}
This factorization can be continuously deformed into an $A_4$-factorization in two different ways.
The first option is,
\begin{eqnarray}
\left(
\begin{array}{cc}
 -x-y & f z^2 +g z\\
 z^4 & x-y \\
\end{array}
\right)
\cdot
\left(
\begin{array}{cc}
 -x+y & f z^2+g z \\
 z^4 & x+y \\
\end{array}
\right)
=(-y^2+x^2+g_0 z^6 + g_1 z^5)\mathbbm{1},
\end{eqnarray}
where $g_{i} \in \mathbbm{C}$ are deformation parameters. The values $(g_0,g_1)=(1,0)$ restore the pure $A_5$ singularity and  $(g_0,g_1)=(0,1)$ the $A_4$ singularity.
The second way is,
\begin{eqnarray}
\left(
\begin{array}{cc}
 -x-y & z^2\\
 fz^4+g z^3 & x-y \\
\end{array}
\right)
\cdot
\left(
\begin{array}{cc}
 -x+y & z^2 \\
 fz^4+g z^3 & x+y \\
\end{array}
\right)
=(-y^2+x^2+g_0 z^6 + g_1 z^5)\mathbbm{1}.
\end{eqnarray}
In a conventional treatment of F-theory the only information about a local 7-brane is the term on the right-hand side of the equation. At the level of matrix factorizations we have a much richer structure: Every node of the Dynkin diagram has a separate description, the open string spectrum between these components of branes can be found by computing the morphisms and, as we have just seen, there can be more than one way to deform a brane. It remains an open question to classify all deformations of the ADE-singularities. It is also an open question which deformations may be ruled out by physical principles. For instance, given a brane located at an $A_n$ singularity, one would expect that the factorization should be a direct sum of all $n$ factorizations (or bound states derived thereof). When deforming these $n$ branes in an arbitrary way to $A_{n-1}$, it is not guaranteed that each of the $n-1$ different $A_{n-1}$ factorizations will be obtained. One could for example deform the factorizations in the following manner:
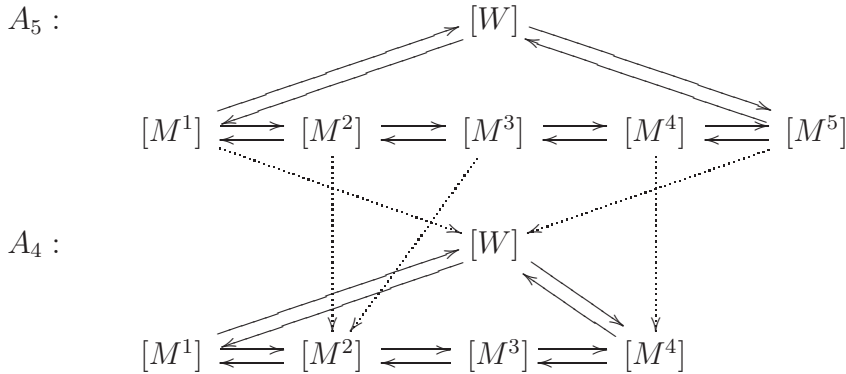
\begin{eqnarray*}
\xymatrix{
A_5:& & &[W]\ar@<0.5ex>[drr]\ar@<0.5ex>[dll] \\
 &[M^1]\; \ar@<0.5ex>[r]\ar@<0.5ex>[urr]\ar@{.>}[drr]
 & \;  [M^2]\ \ar@<0.5ex>[l]\ar@<0.5ex>[r]\ar@{.>}[dd] 
 & \;  [M^3]\ \ar@<0.5ex>[l]\ar@<0.5ex>[r]\ar@{.>}[ddl]
 & \;  [M^4]\ \ar@<0.5ex>[l]\ar@<0.5ex>[r]\ar@{.>}[dd]
 & \;  [M^5]\ \ar@<0.5ex>[l]\ar@<0.5ex>[ull]\ar@{.>}[dll]\\
A_4:& & &[W]\ar@<0.5ex>[dr]\ar@<0.5ex>[dll] \\
& [M^1]\;\ar@<0.5ex>[r]\ar@<0.5ex>[rru]& \;  [M^2]\ \ar@<0.5ex>[l]\ar@<0.5ex>[r]
 & \;  [M^3] \ar@<0.5ex>[l]\ar@<0.5ex>[r]
 & \;  [M^4]\ \ar@<0.5ex>[l]\ar@<0.5ex>[lu]
}
\end{eqnarray*}
In the above diagram the dotted arrows points from the original factorization to the deformed factorization and $[W]$ denotes the trivial factorization which sits at the extended node of the Dynkin diagram. In this example, no brane is deformed into either $M^1$ or $M^3$.
To avoid this situation, we can look at the entire set of all $n$  branes and deform them in the same manner. Then we have either,
\begin{eqnarray*}
E_j=\left(
\begin{array}{cc}
 -x-y & g_0 z^j+g_1 z^{j-1} \\
 z^{n+1-j} & x-y \\
\end{array}
\right)\qquad
J_j=\left(
\begin{array}{cc}
 -x+y & g_0 z^j+g_1 z^{j-1} \\
 z^{n+1-j} & x+y \\
\end{array}
\right)\qquad j=1,...,n,
\end{eqnarray*}
and for $n=5$ get,
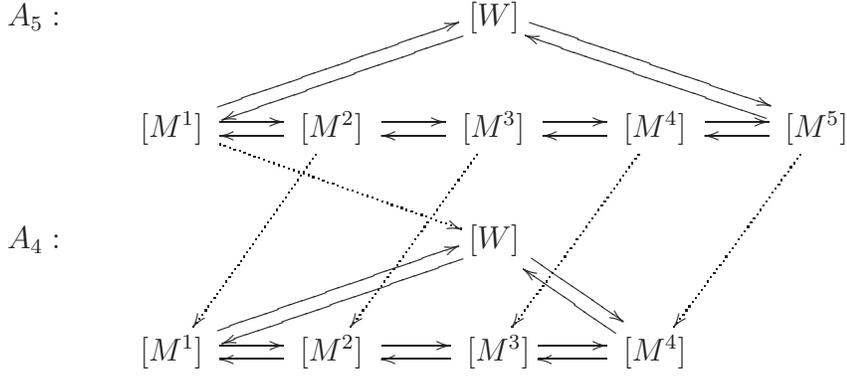
\begin{eqnarray*}
\xymatrix{
A_5:& & &[W]\ar@<0.5ex>[drr]\ar@<0.5ex>[dll] \\
 &[M^1]\; \ar@<0.5ex>[r]\ar@<0.5ex>[urr]\ar@{.>}[drr]
 & \;  [M^2]\ \ar@<0.5ex>[l]\ar@<0.5ex>[r]\ar@{.>}[ddl] 
 & \;  [M^3]\ \ar@<0.5ex>[l]\ar@<0.5ex>[r]\ar@{.>}[ddl]
 & \;  [M^4]\ \ar@<0.5ex>[l]\ar@<0.5ex>[r]\ar@{.>}[ddl]
 & \;  [M^5]\ \ar@<0.5ex>[l]\ar@<0.5ex>[ull]\ar@{.>}[ddl]\\
A_4:& & &[W]\ar@<0.5ex>[dr]\ar@<0.5ex>[dll] \\
& [M^1]\;\ar@<0.5ex>[r]\ar@<0.5ex>[rru]& \;  [M^2]\ \ar@<0.5ex>[l]\ar@<0.5ex>[r]
 & \;  [M^3] \ar@<0.5ex>[l]\ar@<0.5ex>[r]
 & \;  [M^4]\ \ar@<0.5ex>[l]\ar@<0.5ex>[lu]
}
\end{eqnarray*}
Alternatively we deform,
\begin{eqnarray*}
E_j=\left(
\begin{array}{cc}
 -x-y & z^j \\
 g_0 z^{n+1-j}+g_1 z^{n-j} & x-y \\
\end{array}
\right)\qquad
J_j=\left(
\begin{array}{cc}
 -x+y & z^j \\
 g_0 z^{n+1-j}+g_1 z^{n-j} & x+y \\
\end{array}
\right)\qquad j=1,...,n,
\end{eqnarray*}
and obtain,
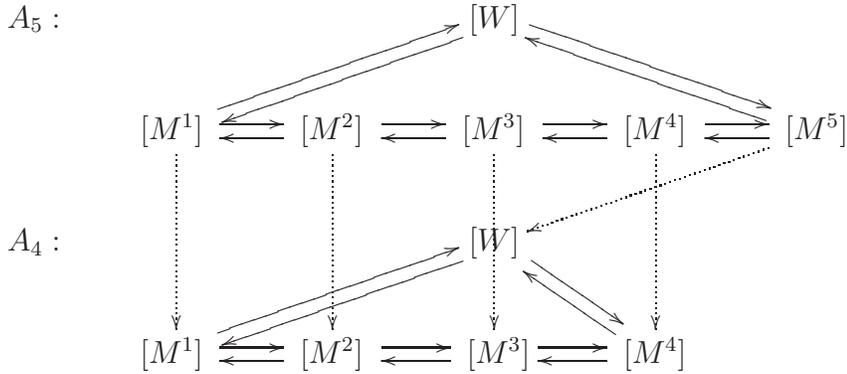
\begin{eqnarray*}
\xymatrix{
A_5:& & &[W]\ar@<0.5ex>[drr]\ar@<0.5ex>[dll] \\
 &[M^1]\; \ar@<0.5ex>[r]\ar@<0.5ex>[urr]\ar@{.>}[dd]
 & \;  [M^2]\ \ar@<0.5ex>[l]\ar@<0.5ex>[r]\ar@{.>}[dd] 
 & \;  [M^3]\ \ar@<0.5ex>[l]\ar@<0.5ex>[r]\ar@{.>}[dd]
 & \;  [M^4]\ \ar@<0.5ex>[l]\ar@<0.5ex>[r]\ar@{.>}[dd]
 & \;  [M^5]\ \ar@<0.5ex>[l]\ar@<0.5ex>[ull]\ar@{.>}[dll]\\
A_4:& & &[W]\ar@<0.5ex>[dr]\ar@<0.5ex>[dll] \\
& [M^1]\;\ar@<0.5ex>[r]\ar@<0.5ex>[rru]& \;  [M^2]\ \ar@<0.5ex>[l]\ar@<0.5ex>[r]
 & \;  [M^3] \ar@<0.5ex>[l]\ar@<0.5ex>[r]
 & \;  [M^4]\ \ar@<0.5ex>[l]\ar@<0.5ex>[lu]
}
\end{eqnarray*}
The latter option is the more conventional choice since the process of deformation here simply removes the $n$-th node of the Dynkin diagram. The former choice is equivalent to the latter after the exchange of branes with anti-branes (which maps $T(M^j)\mapsto M^{n+1-j}$). In the simple case of the $A_n$ singularity we were able to make an argument of deforming all branes consistently and relied on the fact that all branes have the same factorization structure. With other types of singularities things are not as straightforward and it remains unclear which of the possible deformations are preferred.
\section{Deformation from $E_8$ to $D_5$}
As stated in the introduction, one purpose of this paper is to move beyond branes for simple toy models and demonstrate that branes which appear in phenomenologically viable models can be described. In this section I will take all indecomposable matrix factorizations and deform them in the sequence $E_8 \rightarrow E_7 \rightarrow E_6 \rightarrow D_5$. The $A_n$ factorizations are rather simple and the further deformation to $A_4$ is not worked out here. Given the singularity of Eq.~(\ref{eq:W}),
\begin{eqnarray*}
W(x,y,z)=-y^2+f_1 x^3+f_2 x^2 z+f_3^2 x^2+g_1 z^5+g_2 x z^3+g_3^2 z^4,
\end{eqnarray*}
we can reproduce the $E_{8,7,6}$ and $D_5$ singularities by setting the appropriate
coefficients of $f_i$ and $g_i$ to zero. In principle the $f_3$-term is not necessary for the breaking pattern, but it is useful to preserve it for the straightforward extension to the $A_n$ singularities. For a more compact notation of the factorizations we define for later use,
\begin{eqnarray*}
\begin{array}{rcl}
F&:=&f_1 x + f_2 z  + f_3^2\\
G&:=&g_1 z^2 + g_2 x + g_3^2 z
\end{array}
\end{eqnarray*}
\subsection{Brane $M^1$}
A deformation of the brane $M^1$ of $E_8$ is given by,
\begin{eqnarray*}
\tilde{M}^1:(\tilde{S}_1-y\mathbbm{1},\tilde{S}_1+y\mathbbm{1})
\end{eqnarray*}
where,
\begin{eqnarray*}
\tilde{S}_1:=\left(
\begin{array}{cccc}
 0 & 0 & F x & G z^2 \\
 0 & 0 & z & -x \\
 x & G z^2 & 0 & 0 \\
 z & -Fx & 0 & 0 \\
\end{array}
\right).
\end{eqnarray*}
By direct computation it is easy to verify that the factorization condition is satified,
\begin{eqnarray*}
(\tilde{S}_1-y\mathbbm{1})\cdot(\tilde{S}_1+y\mathbbm{1})=
(-y^2+f_1 x^3+f_2 x^2 z+f_3^{2} x^2+g_1 z^5+g_2 x z^3+g_3^2 z^4)\mathbbm{1}.
\end{eqnarray*}
The deformation sequence for this brane is rather simple:
\begin{eqnarray*}
\xymatrix{
E_8: &[M^1]\; \ar@{.>}[d]\\
E_7: &[M^1]\; \ar@{.>}[d]\\
E_6: &[M^1]\; \ar@{.>}[d]\\
D_5: &[M^1]\; 
}
\end{eqnarray*}
This can be seen by setting the appropriate coefficients $f_i$ and $g_i$ to zero respectively unity and comparing with the factorization list of the ADE-singularities.
\subsection{Brane $M^2$}
A deformation of the brane $M^2$ of $E_8$ is given by,
\begin{eqnarray*}
\tilde{M}_2:(\tilde{S}_2-y\mathbbm{1},\tilde{S}_2+y\mathbbm{1})
\end{eqnarray*}
where,
\begin{eqnarray*}
\tilde{S}_2:=\left(
\begin{array}{cccccc}
 0 & 0 & 0 & F x & G z^2 & -G x z \\
 0 & 0 & 0 & -F z & F x & G z^2 \\
 0 & 0 & 0 & z^2 & -x z & x^2 \\
 x & 0 & G z & 0 & 0 & 0 \\
 z & x & 0 & 0 & 0 & 0 \\
 0 & z & F & 0 & 0 & 0 \\
\end{array}
\right)
\end{eqnarray*}
The deformation sequence of this brane is:
\begin{eqnarray*}
\xymatrix{
E_8: &[M^2]\; \ar@{.>}[d]\\
E_7: &[M^2]\; \ar@{.>}[d]\\
E_6: &[M^2]\; \ar@{.>}[d]\\
D_5: &[M^1] \oplus [M^3]\; 
}
\end{eqnarray*}
Again it is manifest from the list of factorizations how the $E_8$ factorizations deforms down to $E_6$, but the last step is non-trivial. For $f_1=f_3=g_1=g_2=0$ we obtain a deformation to $D_5$ which reads,
\begin{eqnarray*}
\tilde{M}^2(D_5):(\tilde{S}_2(D_5)-y\mathbbm{1},\tilde{S}_2(D_5)+y\mathbbm{1})
\end{eqnarray*}
where,
\begin{eqnarray*}
\tilde{S}_2(D_5):=\left(
\begin{array}{cccccc}
 0 & 0 & 0 & f_2 x z & g_3^2 z^3 & -g_3^2 x z^2 \\
 0 & 0 & 0 & -f_2z^2 & f_2 x z & g_3^2 z^3 \\
 0 & 0 & 0 & z^2 & -x z & x^2 \\
 x & 0 & g_3^2 z^2 & 0 & 0 & 0 \\
 z & x & 0 & 0 & 0 & 0 \\
 0 & z & f_2 z & 0 & 0 & 0 \\
\end{array}
\right)
\end{eqnarray*}
After an appropriate gauge transformation, this factorization decomposes into a direct sum of smaller matrices. To realize the gauge transformation we define the matrix,
\begin{eqnarray*}
U:=\left(
\begin{array}{cccccc}
 0 & 1 & f_2 & 0 & 0 & 0 \\
 0 & 0 & 0 & 0 & 0 & 1 \\
 1 & 0 & 0 & 0 & 0 & 0 \\
 0 & -1 & 0 & 0 & 0 & 0 \\
 0 & 0 & 0 & f_2 & 0 & -g_3^2z \\
 0 & 0 & 0 & 0 & 1 & 0 \\
\end{array}
\right)
\end{eqnarray*}
We can assume $f_2$ and $g_3$ to be locally non-zero, therefore $U$ is invertible as required for a well-defined similarity transformation.
Then,
\begin{eqnarray*}
U\cdot(\tilde{S}_2(D_5)\pm y\mathbbm{1})\cdot U^{-1}=
\left(
\begin{array}{cccccc}
 \pm y & g_3^2 z^3+f_2 x^2  & 0 & 0 & 0 & 0 \\
 z & \pm y & 0 & 0 & 0 & 0 \\
 0 & 0 & \pm y & 0 & x z & g_3^2z^3  \\
 0 & 0 & 0 & \pm y & z^2 & -f_2 x z  \\
 0 & 0 & f_2 x & g_3^2 z^2 & \pm y & 0 \\
 0 & 0 & z & -x & 0 & \pm y \\
\end{array}
\right)
\end{eqnarray*}
For $f_2=g_3=1$ the two block matrices on the right-hand side are identified with $M^1(D_5)$ and $M^3(D_5)$ in the factorization list of the simple singularities.
\subsection{Brane $M^3$}
A deformation of the brane $M^3$ of $E_8$ is given by,
\begin{eqnarray*}
\tilde{M}^3:(\tilde{S}_3-y\mathbbm{1},\tilde{S}_3+y\mathbbm{1})
\end{eqnarray*}
where,
\begin{eqnarray*}
\tilde{S}_3:=\left(
\begin{array}{cccccccc}
 0 & 0 & 0 & 0 & x & G z & 0 & z \\
 0 & 0 & 0 & 0 & z^2 & -F x & -F z & 0 \\
 0 & 0 & 0 & 0 & 0 & 0 & F x & z^2 \\
 0 & 0 & 0 & 0 & 0 & 0 & G z & -x \\
 F x & G z & 0 & F z & 0 & 0 & 0 & 0 \\
 z^2 & -x & -z & 0 & 0 & 0 & 0 & 0 \\
 0 & 0 & x & z^2 & 0 & 0 & 0 & 0 \\
 0 & 0 & G z & -F x & 0 & 0 & 0 & 0 \\
\end{array}
\right)
\end{eqnarray*}
The deformation sequence is:
\begin{eqnarray*}
\xymatrix{
E_8:  &[M^3]\; \ar@{.>}[d] \\
E_7: &[M^3]\; \ar@{.>}[d] \\
E_6: & [M^3]\oplus [M^4]\; \ar@{.>}[d] \\
D_5: &[M^3] \oplus [M^3] 
}
\end{eqnarray*}
The decomposition at the $E_6$ level is proven by the gauge transformation,
\begin{eqnarray*}
\begin{array}{l}
U_1 \cdot \left(
\begin{array}{cccccccc}
 -y & 0 & 0 & 0 & x & g_3^2 z^2 & 0 & z \\
 0 & -y & 0 & 0 & z^2 & -f_1 x^2 & -f_1 x z & 0 \\
 0 & 0 & -y & 0 & 0 & 0 & f_1 x^2 & z^2 \\
 0 & 0 & 0 & -y & 0 & 0 & g_3^2 z^2 & -x \\
 f_1 x^2 & g_3^2 z^2 & 0 & f_1 x z & -y & 0 & 0 & 0 \\
 z^2 & -x & -z & 0 & 0 & -y & 0 & 0 \\
 0 & 0 & x & z^2 & 0 & 0 & -y & 0 \\
 0 & 0 & g_3^2 z^2 & -f_1 x^2 & 0 & 0 & 0 & -y \\
\end{array}
\right)\cdot U_2= \\
\left(
\begin{array}{cccccccc}
 i y-i g_3 z^2 & f_1 x^2 & 0 & g_3 x z & 0 & 0 & 0 & 0 \\
 -x & i y+i g_3 z^2 & g_3 z & 0 & 0 & 0 & 0 & 0 \\
 0 & 0 & i y-i g_3 z^2 & x^2 & 0 & 0 & 0 & 0 \\
 0 & 0 & -f_1 x & i y+i g_3 z^2 & 0 & 0 & 0 & 0 \\
 0 & 0 & 0 & 0 & i y+i g_3 z^2 & -f_1 x^2 & 0 & -g_3 x z \\
 0 & 0 & 0 & 0 & x & i y-i g_3 z^2 & -g_3 z & 0 \\
 0 & 0 & 0 & 0 & 0 & 0 & i y+i g_3 z^2 & -x^2 \\
 0 & 0 & 0 & 0 & 0 & 0 & f_1 x & i y-ig_3 z^2 \\
\end{array}
\right)
\end{array}
\end{eqnarray*}
with,
\begin{eqnarray*}
\begin{array}{l}
U_1 = \left(
\begin{array}{cccccccc}
 0 & g_3 & 0 & 0 & 1 & 0 & 0 & 0 \\
 -i & 0 & 0 & 0 & 0 & i g_3 & 0 & 0 \\
 0 & 0 & -1 & 0 & 0 & 0 & 0 & -\frac{1}{g_3} \\
 0 & 0 & 0 & -\frac{i f_1}{g_3} & 0 & 0 & i f_1 & 0 \\
 0 & -i g_3 & 0 & 0 & i & 0 & 0 & 0 \\
 -1 & 0 & 0 & 0 & 0 & -g_3 & 0 & 0 \\
 0 & 0 & i & 0 & 0 & 0 & 0 & -\frac{i}{g_3} \\
 0 & 0 & 0 & -\frac{f_1}{g_3} & 0 & 0 & -f_1 & 0 \\
\end{array}
\right)\\\\
U_2=\left(
\begin{array}{cccccccc}
 0 & \frac{1}{2} & 0 & 0 & 0 & \frac{i}{2} & 0 & 0 \\
 -\frac{i}{2 g_3} & 0 & 0 & 0 & \frac{1}{2 g_3} & 0 & 0 & 0 \\
 0 & 0 & \frac{i}{2} & 0 & 0 & 0 & -\frac{1}{2} & 0 \\
 0 & 0 & 0 & \frac{g_3}{2 f_1} & 0 & 0 & 0 & \frac{i g_3}{2 f_1} \\
 -\frac{i}{2} & 0 & 0 & 0 & -\frac{1}{2} & 0 & 0 & 0 \\
 0 & -\frac{1}{2 g_3} & 0 & 0 & 0 & \frac{i}{2 g_3} & 0 & 0 \\
 0 & 0 & 0 & -\frac{1}{2 f_1} & 0 & 0 & 0 & \frac{i}{2 f_1} \\
 0 & 0 & \frac{i g_3}{2} & 0 & 0 & 0 & \frac{g_3}{2} & 0 \\
\end{array}
\right)
\end{array}
\end{eqnarray*}
At the $D_5$-level a similar transformation exists which is not explicitly written down here. Remember that $M^3(E_6)$
and $M^4(E_6)$ are brane/anti-brane pairs,
\begin{eqnarray}
T(M_3(E_6))=M_4(E_6).
\end{eqnarray}
On the other hand, $M_3(D_5)$ is its own anti-brane, i.e. on the level 
of factorizations it is self-dual,
\begin{eqnarray}
T(M_3(D_5))=M_3(D_5)
\end{eqnarray}
Just from these relations it is clear that the deformation from $M_3(E_6)$ to $M_3(D_5)$ implies that $M_4(E_6)$ deforms also into $M_3(D_5)$.
Therefore one brane at the $E_8$ level has been deformed into a direct sum of two identical branes. However, we expect the physics of a system with a direct sum of identical branes not to differ from the physics of a system with only single copy of the factorization.
\subsection{Brane $M^4$}
A deformation of the brane $M^4$ of $E_8$ is given by,
\begin{eqnarray*}
\tilde{M}^4:(\tilde{S}_4-y\mathbbm{1},\tilde{S}_4+y\mathbbm{1})
\end{eqnarray*}
where,
\begin{eqnarray*}
\tilde{S}_4:=\left(
\begin{array}{cccccccccc}
 0 & 0 & 0 & 0 & 0 & F x & z^2 & z^3 & 0 & -F z \\
 0 & 0 & 0 & 0 & 0 & g z & -x & 0 & -z^2 & 0 \\
 0 & 0 & 0 & 0 & 0 & 0 & 0 & x^2 & -x z & g z \\
 0 & 0 & 0 & 0 & 0 & 0 & 0 & g z^2 & F x & -F g \\
 0 & 0 & 0 & 0 & 0 & 0 & 0 & -x z^2 & z^3 & F x \\
 x & z^2 & 0 & 0 & z & 0 & 0 & 0 & 0 & 0 \\
 g z & -F x & 0 & -z^2 & 0 & 0 & 0 & 0 & 0 & 0 \\
 0 & 0 & F & z & 0 & 0 & 0 & 0 & 0 & 0 \\
 0 & 0 & 0 & x & g & 0 & 0 & 0 & 0 & 0 \\
 0 & 0 & z^2 & 0 & x & 0 & 0 & 0 & 0 & 0 \\
\end{array}
\right)
\end{eqnarray*}
The deformation sequence is:
\begin{eqnarray*}
\xymatrix{
E_8: & [M^4]\; \ar@{.>}[d] \\
E_7: & [M^4] \oplus [M^5]\; \ar@{.>}[d] \\
E_6: & [M^2] \oplus [M^5] \oplus [M^6]\; \ar@{.>}[d] \\
D_5: & [M^1] \oplus [M^2] \oplus [M^4] \oplus [M^5] 
}
\end{eqnarray*}
\subsection{Brane $M^5$}
A deformation of the brane $M^5$ of $E_8$ is given by,
\begin{eqnarray*}
\tilde{M}^5:(\tilde{S}_5-y\mathbbm{1},\tilde{S}_5+y\mathbbm{1})
\end{eqnarray*}
where,
\begin{eqnarray*}
\tilde{S}_5:=\left(
\begin{array}{cccccccccccc}
 0 & 0 & 0 & 0 & 0 & 0 & F x & -F z & G z & 0 & F & 0 \\
 0 & 0 & 0 & 0 & 0 & 0 & G z^2 & F x & -G x & 0 & 0 & x z \\
 0 & 0 & 0 & 0 & 0 & 0 & -x z^2 & z^3 & x^2 & x z & 0 & 0 \\
 0 & 0 & 0 & 0 & 0 & 0 & 0 & 0 & 0 & x & z & 0 \\
 0 & 0 & 0 & 0 & 0 & 0 & 0 & 0 & 0 & 0 & F & z^2 \\
 0 & 0 & 0 & 0 & 0 & 0 & 0 & 0 & 0 & G & 0 & x \\
 x & z & 0 & 0 & -x & 0 & 0 & 0 & 0 & 0 & 0 & 0 \\
 0 & x & G & 0 & 0 & -x z & 0 & 0 & 0 & 0 & 0 & 0 \\
 z^2 & 0 & F & -F z & 0 & 0 & 0 & 0 & 0 & 0 & 0 & 0 \\
 0 & 0 & 0 & F x & -x z & z^3 & 0 & 0 & 0 & 0 & 0 & 0 \\
 0 & 0 & 0 & G z^2 & x^2 & -x z^2 & 0 & 0 & 0 & 0 & 0 & 0 \\
 0 & 0 & 0 & -F G & G z & F x & 0 & 0 & 0 & 0 & 0 & 0 \\
\end{array}
\right)
\end{eqnarray*}
The deformation sequence is:
\begin{eqnarray*}
\xymatrix{
E_8: & [M^5]\; \ar@{.>}[d] \\
E_7: & [M^3] \oplus [M^7] \oplus [M^7]\; \ar@{.>}[d] \\
E_6: & [M^2] \oplus [M^2]\; \ar@{.>}[d] \\
D_5: & [M^1] \oplus [M^1] \oplus [M^3] \oplus [M^3] 
}
\end{eqnarray*}
In the deformation sequences one could most of the times specify which of the factorizations deform to which lower rank factorizations, for instance each $M^2(E_6)$ deforms into one copy of $M^1(D_5) \oplus M^3(D_5)$, but this does not hold every time. For example at the $E_7$-level of this brane, a sum of branes is required to deform to the $E_6$ factorization.
\subsection{Brane $M^6$}
A deformation of the brane $M^6$ of $E_8$ is given by,
\begin{eqnarray*}
\tilde{M}^6:(\tilde{S}_6-y\mathbbm{1},\tilde{S}_6+y\mathbbm{1})
\end{eqnarray*}
where,
\begin{eqnarray*}
\tilde{S}_6:=\left(
\begin{array}{cccccc}
 0 & 0 & 0 & F x & -F z & G z \\
 0 & 0 & 0 & G z^2 & F x & -G x \\
 0 & 0 & 0 & -x z^2 & z^3 & x^2 \\
 x & z & 0 & 0 & 0 & 0 \\
 0 & x & G & 0 & 0 & 0 \\
 z^2 & 0 & F & 0 & 0 & 0 \\
\end{array}
\right)
\end{eqnarray*}
The deformation sequence is:
\begin{eqnarray*}
\xymatrix{
E_8: & [M^6]\; \ar@{.>}[d] \\
E_7: & [M^4] \oplus [M^7]\; \ar@{.>}[d] \\
E_6: & [M^2]\; \ar@{.>}[d] \\
D_5: & [M^1]\oplus [M^3] 
}
\end{eqnarray*}
Note that the $E_8$ factorization first falls apart into a direct sum of two $E_7$ factorizations but at further deformation to $E_6$ the two components recombine into a single one. The fact that deformations can involve and sometimes has to involve a direct sum of factorizations rather than being restricted to single factorization only makes a systematic treatment of all possible deformations much more difficult. 
\subsection{Brane $M^7$}
A deformation of the brane $M^7$ of $E_8$ is given by,
\begin{eqnarray*}
\tilde{M}^7:(\tilde{S}_7-y\mathbbm{1},\tilde{S}_7+y\mathbbm{1})
\end{eqnarray*}
where,
\begin{eqnarray*}
\tilde{S}_7:=\left(
\begin{array}{cccccccc}
 0 & 0 & 0 & 0 & x & z^2 & 0 & -z \\
 0 & 0 & 0 & 0 & G z & -F x & G x & 0 \\
 0 & 0 & 0 & 0 & 0 & 0 & x^2 & z^2 \\
 0 & 0 & 0 & 0 & 0 & 0 & G z & -F \\
 F x & z^2 & 0 & -x z & 0 & 0 & 0 & 0 \\
 G z & -x & G & 0 & 0 & 0 & 0 & 0 \\
 0 & 0 & F & z^2 & 0 & 0 & 0 & 0 \\
 0 & 0 & G z & -x^2 & 0 & 0 & 0 & 0 \\
\end{array}
\right)
\end{eqnarray*}
The deformation sequence is:
\begin{eqnarray*}
\xymatrix{
E_8: & [M^7]\; \ar@{.>}[d] \\
E_7: & [M^5] \oplus [M^7]\; \ar@{.>}[d] \\
E_6: & [M^3] \oplus [M^4]\; \ar@{.>}[d] \\
D_5: & [M^1]\oplus [M^1]\oplus [M^4] \oplus [M^5]
}
\end{eqnarray*}
\subsection{Brane $M^8$}
A deformation of the brane $M^8$ of $E_8$ is given by,
\begin{eqnarray*}
\tilde{M}^8:(\tilde{S}_8-y\mathbbm{1},\tilde{S}_8+y\mathbbm{1})
\end{eqnarray*}
where,
\begin{eqnarray*}
\tilde{S}_8:=\left(
\begin{array}{cccc}
 0 & 0 & F x & G z \\
 0 & 0 & z^2 & -x \\
 x & G z & 0 & 0 \\
 z^2 & -F x & 0 & 0 \\
\end{array}
\right)
\end{eqnarray*}
The deformation sequence is:
\begin{eqnarray*}
\xymatrix{
E_8: & [M^8]\; \ar@{.>}[d] \\
E_7: & [M^6] \; \ar@{.>}[d] \\
E_6: & [M^5] \oplus [M^6]\; \ar@{.>}[d] \\
D_5: & [M^4]\oplus [M^5]
}
\end{eqnarray*}
Given the small matrix dimensions of this example, it is worth looking at it explicitly.
\begin{eqnarray*}
E_8: \left(
\begin{array}{cccc}
 \pm y & 0 & f_1 x^2 & g_1 z^3 \\
 0 & \pm y & z^2 & -x \\
 x & g_1 z^3 & \pm y & 0 \\
 z^2 & -f_1 x^2 & 0 & \pm y \\
\end{array}
\right)
\end{eqnarray*}
\begin{eqnarray*}
E_7: \left(
\begin{array}{cccc}
 \pm y & 0 & f_1 x^2 & g_2 x z \\
 0 & \pm y & z^2 & -x \\
 x & g_2 x z & \pm y & 0 \\
 z^2 & -f_1 x^2 & 0 & \pm y \\
\end{array}
\right)
\end{eqnarray*}
\begin{eqnarray*}
E_6: U_a\left(
\begin{array}{cccc}
 \pm y & 0 & f_1 x^2 & g_3^2 z^2 \\
 0 & \pm y & z^2 & -x \\
 x & g_3^2 z^2 & \pm y & 0 \\
 z^2 & -f_1 x^2 & 0 & \pm y \\
\end{array}
\right)U_a^{-1}=\left(
\begin{array}{cccc}
 \pm y+g_3 z^2 & f_1 x^2 & 0 & 0 \\
 x & \pm y-g_3 z^2 & 0 & 0 \\
 0 & 0 &\pm  y-g_3 z^2 & f_1 x^2 \\
 0 & 0 & x & \pm y +g_3 z^2 \\
\end{array}
\right)
\end{eqnarray*}
\begin{eqnarray*}
D_5: U_a\left(
\begin{array}{cccc}
 \pm y & 0 & f_2 xz & g_3^2 z^2 \\
 0 & \pm y & z^2 & -x \\
 x & g_3^2 z^2 & \pm y & 0 \\
 z^2 & -f_2 x z & 0 & \pm y \\
\end{array}
\right)U_a^{-1}=\left(
\begin{array}{cccc}
 \pm y + g_3 z^2  & f_2 x z & 0 & 0 \\
 x & \pm y-g_3 z^2 & 0 & 0 \\
 0 & 0 & \pm y -g_3 z^2 & f_2 x z \\
 0 & 0 & x & \pm y +g_3z^2  \\
\end{array}
\right)
\end{eqnarray*}
\begin{eqnarray*}
A_4: U_b\left(
\begin{array}{cccc}
 \pm y & 0 & f_3^2 x & g_1 z^3 \\
 0 & \pm y & z^2 & -x \\
 x & g_1 z^3 & \pm y & 0 \\
 z^2 & -f_3^2 x & 0 & \pm y \\
\end{array}
\right)U_b^{-1}=\left(
\begin{array}{cccc}
 \pm y-f_3 x & z^2 & 0 & 0 \\
 g_1 z^3 &\pm y+ f_3 x & 0 & 0 \\
 0 & 0 & \pm y-f_3 x & g_1 z^3 \\
 0 & 0 & z^2 & \pm y + f_3 x \\
\end{array}
\right)
\end{eqnarray*}
The two transformation matrices which were used to turn the matrices into a sum of indecomposable factorizations are given by,
\begin{eqnarray*}
U_{a}=\left(
\begin{array}{cccc}
 1 & 0 & 0 & g_3 \\
 0 & -g_3 & 1 & 0 \\
 1 & 0 & 0 & -g_3 \\
 0 & g_3 & 1 & 0 \\
\end{array}
\right)
\qquad
U_{b}=\left(
\begin{array}{cccc}
 0 & f_3 & 0 & 1 \\
 1 & 0 & f_3 & 0 \\
 1 & 0 & -f_3 & 0 \\
 0 & -f_3 & 0 & 1 
\end{array}
\right).
\end{eqnarray*}
\bibliographystyle{JHEP}
\bibliography{ftheory,matfac}                            
\end{document}